\documentclass[conference]{IEEEtran}
\IEEEoverridecommandlockouts
\usepackage{cite}
\usepackage{amsmath,amssymb,amsfonts}
\usepackage{algorithmic}
\usepackage{graphicx}
\usepackage{textcomp}
\usepackage{xcolor}
\def\BibTeX{{\rm B\kern-.05em{\sc i\kern-.025em b}\kern-.08em
    T\kern-.1667em\lower.7ex\hbox{E}\kern-.125emX}}
    
\usepackage{url}
\usepackage[caption=false,font=footnotesize]{subfig}
    
\begin{document}


\title{Training Data Improvement for \\ Image Forgery Detection using Comprint\\
{\footnotesize
\thanks{This work was funded in part by the Research Foundation -- Flanders (FWO), IDLab (Ghent University -- imec), Flanders Innovation \& Entrepreneurship (VLAIO), the Flemish Government, and the European Union. In addition, this material is based on research sponsored by the Defense Advanced Research Projects Agency (DARPA) and the Air Force Research Laboratory (AFRL) under agreement number FA8750-20-2-1004. }
}}

\author{\IEEEauthorblockN{Hannes Mareen$^1$, Dante Vanden Bussche$^1$, Glenn Van Wallendael$^1$, Luisa Verdoliva$^2$ and Peter Lambert$^1$}
\IEEEauthorblockA{
$^1$ Ghent University \--- imec, IDLab, Department of Electronics and Information Systems, Ghent, Belgium, \\
\{hannes.mareen, dante.vandenbussche, glenn.vanwallendael, peter.lambert\}@ugent.be, \url{https://media.idlab.ugent.be} \\
$^2$ Università degli Studi di Napoli Federico II, Naples, Italy, verdoliv@unina.it, \url{https://www.grip.unina.it}
}
}

\maketitle

\begin{abstract}
Manipulated images are a threat to consumers worldwide, when they are used to spread disinformation. Therefore, Comprint enables forgery detection by utilizing JPEG-compression fingerprints. This paper evaluates the impact of the training set on Comprint's performance. Most interestingly, we found that including images compressed with low quality factors during training does not have a significant effect on the accuracy, whereas incorporating recompression boosts the robustness. As such, consumers can use Comprint on their smartphones to verify the authenticity of images.
\end{abstract}

\begin{IEEEkeywords}
Image Forensics, Forgery Detection, Forgery Localization, Deep Learning
\end{IEEEkeywords}

\section{Introduction}
Images are spread at a rapid pace on social media, without guarantees on their authenticity. Yet, such images could have been manipulated using editing tools such as Adobe Photoshop or recent AI-based software. To detect such manipulations, forgery detection methods were developed~\cite{verdoliva2020overview, alZahir2017blind, lee2007detecting}.

A recent, promising method is Comprint~\cite{mareen2020comprint}, which utilizes compression fingerprints to detect forgeries in images. The underlying assumption is that the forged area underwent different compression than the real area, thus generating another compression fingerprint. In other words, inconsistencies in the compression traces of an image suggest manipulation.

This paper evaluates the impact of the training data used to create a comprint. For the purpose of this short paper, we limited the analysis to verifying if using a larger dataset with lower JPEG quality factors (QFs) is beneficial, and if incorporating recompression examples during training increases the robustness against recompression.

\section{Image Forgery Detection using Comprint}
To detect forgeries, the Comprint-algorithm transforms an image to a compression fingerprint or \emph{comprint}. Then, the comprint is transformed to a heatmap which clusters the comprint in two regions (i.e., a real and fake region). 

We train a Convolutional Neural Network (CNN) to transform an image to a comprint. First, the CNN is pre-trained to estimate the JPEG compression artifacts in an image.
Then, we further apply Siamese training~\cite{cozzolino2020noiseprint}. That is, the training is performed per pair of images: each pair either underwent the same or different compression. During training, the distance between two comprints of a pair that underwent the \emph{same} compression should be small (i.e., they ideally create the same comprint). In contrast, a pair with \emph{different} compression should have a large corresponding distance.

After the CNN is trained, the extracted comprint is transformed to a heatmap that localizes potential forgeries. This is done by first extracting co-occurrence-based features~\cite{pevny2010spam}, and then feeding these multi-dimensional features to an Expectation-Maximization algorithm~\cite{cozzolino2015splicebuster}. In this way, each pixel of the image is assigned a continuous number which represents the likelihood of it belonging to either the forged or pristine region. In other words, a heatmap is created that can be used for forgery localization.

\section{Evaluation: Training Data Improvement}
We evaluate the impact of changing the training data on Comprint.
The training, validation and test images are obtained from the RAISE dataset~\cite{dangnguyen2015raise}. From this dataset, 1000 images were randomly selected for training, another 100 for validation, and 50 for testing.
The training and validation images were converted to grayscale, resized to 200x200 pixels, JPEG compressed with certain QFs, and optionally recompressed.
To evaluate the effect of using more QFs and recompression, we created three versions of the training dataset, resulting in three models:

\begin{itemize}
  \item \textbf{HighQF}: QFs in \{50, 55, 60, 65, 70, 80, 90\}.
  \item \textbf{WideQF}: QFs in \{20, 25, 30, 35, 40, 50, 60, 70, 80, 90\}. In other words, it includes lower QF values than HighQF.
  \item \textbf{HighQFRec}: Same QFs as in HighQF (\{50, 55, 60, 65, 70, 80, 90\}), as well as recompression with Recompression QFs (Rec. QFs) in \{50, 55, 60, 65, 70, 80, 90\}. Recompression is performed with 50\% probability.
\end{itemize}

The test images were converted to grayscale and resized to 1000x1000 pixels. Then, we created so-called composite images consisting of two halves: the left half was compressed using a QF in the set \{20, 25, 30, 35, 40, 45, 50, 55, 60, 65, 70, 75, 80, 85, 90\}, and the right half with a QF that is 10 higher than the left half. Then, the resulting image is both saved as PNG (i.e., lossless compression), and recompressed using JPEG with QFs \{50, 60, 70, 80, 90, 95, 100\}. As such, one half should be detected as forged and the other as pristine.

\begin{figure}[!t]
  \centering
  \includegraphics[width=1.0\linewidth]{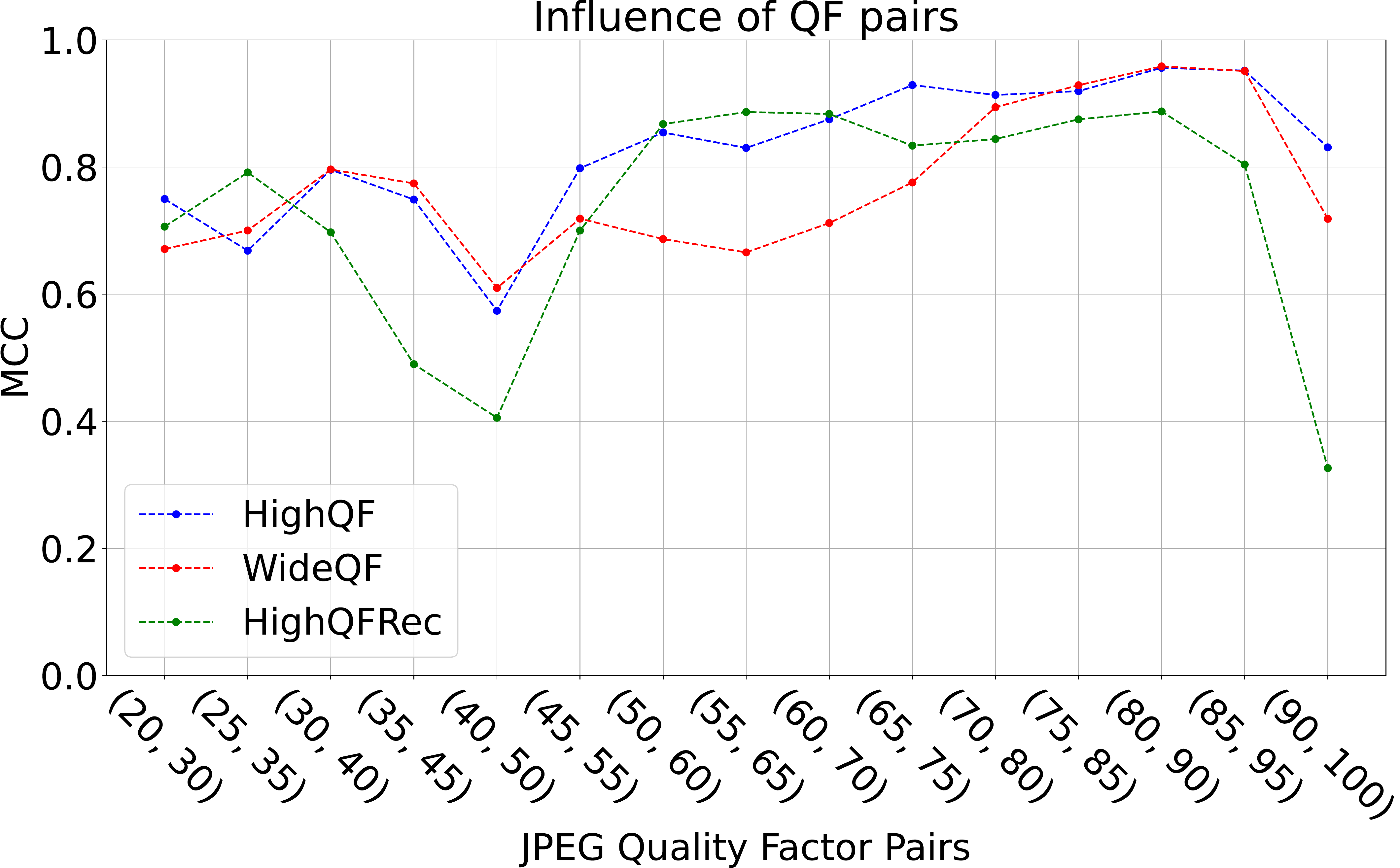}
  \caption{The performance of the three evaluated models on composited images.}
  \vspace{-0.4cm}
  \label{fig:evaluation-diff10}
\end{figure}

\begin{figure}[!t]
\subfloat[HighQF~~~~~~~~~~~~~~~~~~~~~~~~~~~~~~]{\includegraphics[width=\linewidth]{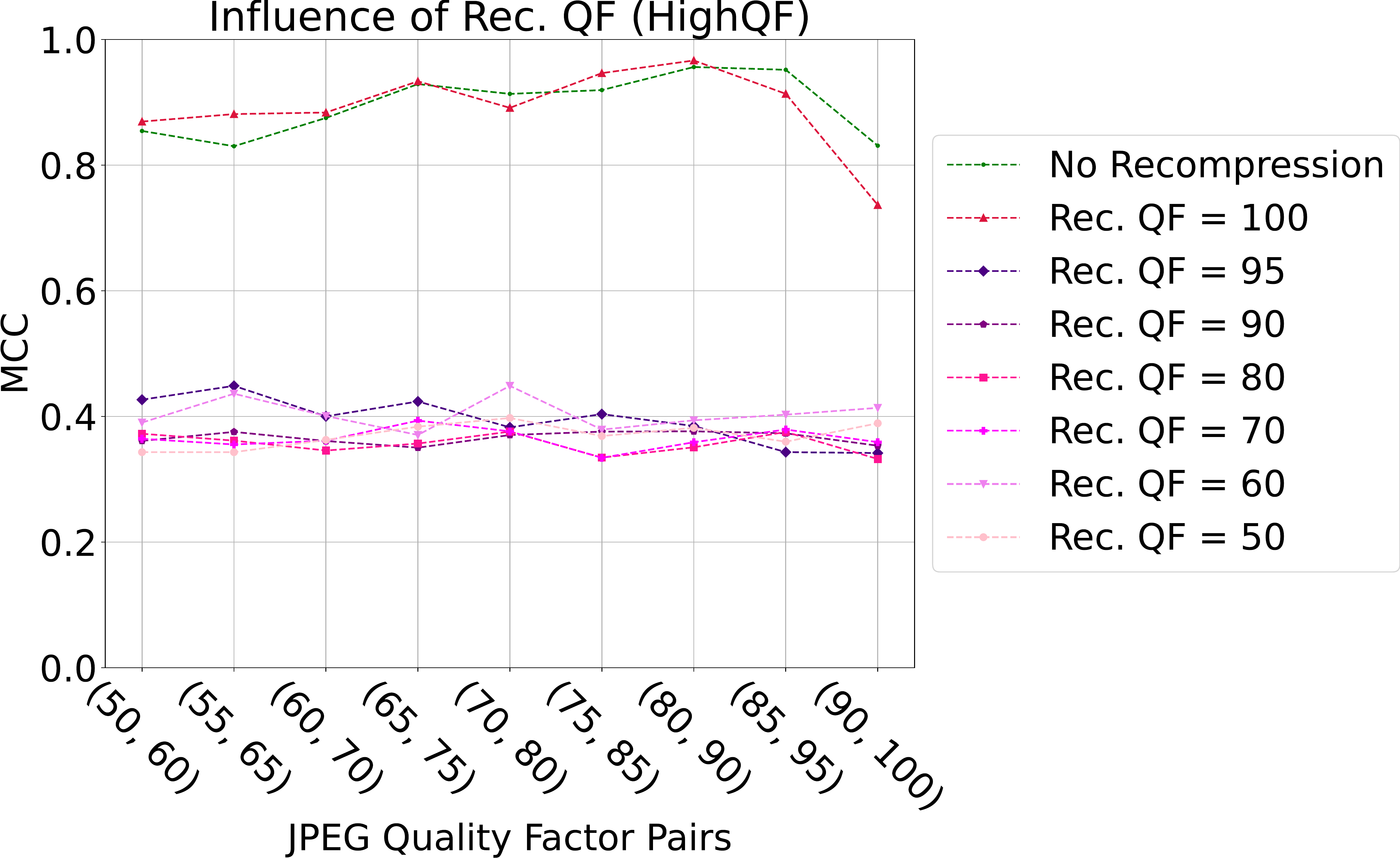}
\label{fig:evaluation-recompress-high}}

\subfloat[HighQFRec]{\includegraphics[width=\linewidth]{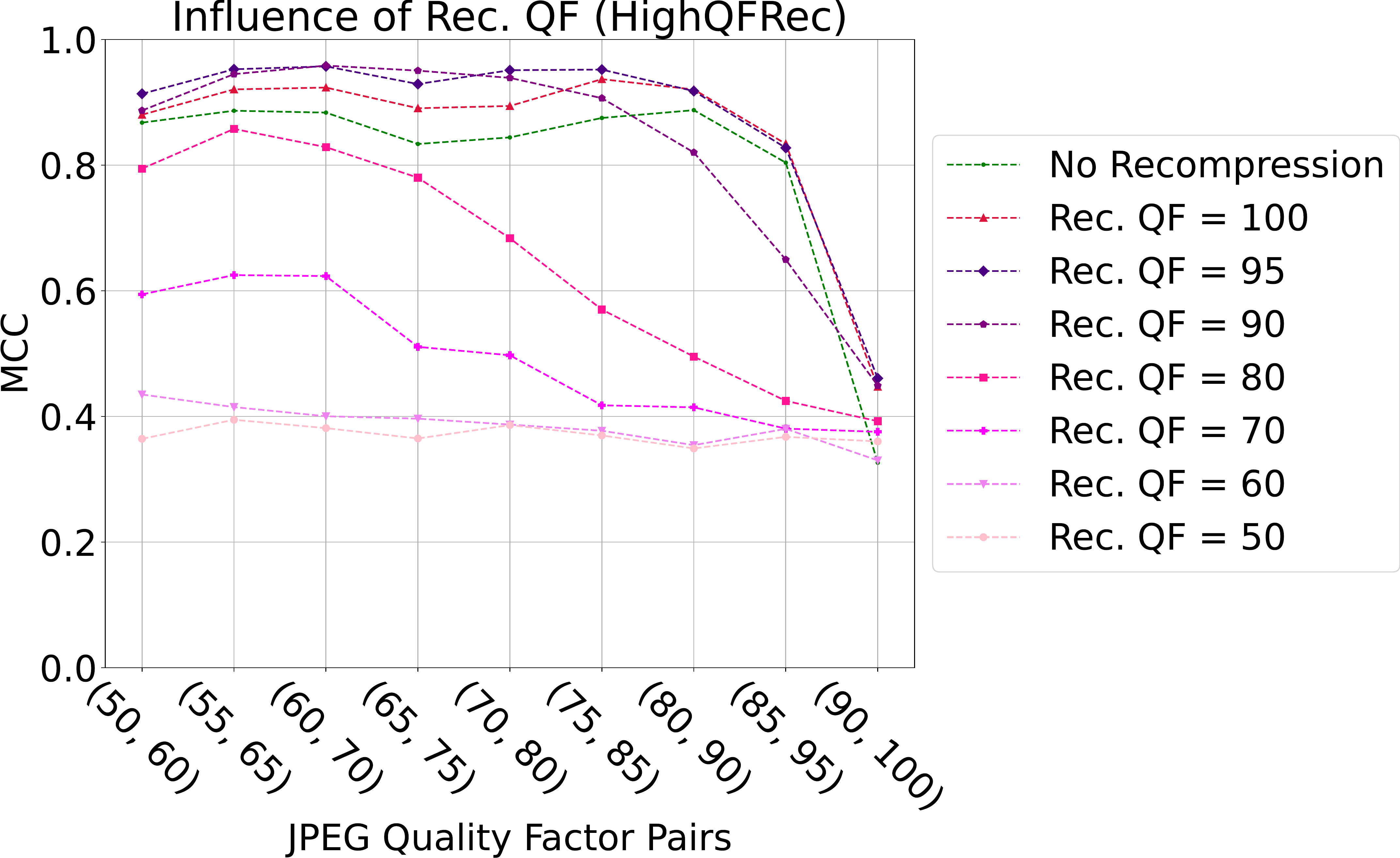}
\label{fig:evaluation-recompress-rec}}
\caption{Performance on recompressed dataset. (a) HighQF is not robust to recompression, whereas (b) HighQFRec is robust to up to a Rec. QF of 60.}
\vspace{-0.6cm}
\label{fig:evaluation-recompress}
\end{figure}

We evaluate the accuracy using Matthew's Correlation Coefficient (MCC). The MCC is defined in (\ref{eq:mcc}), and characterizes the correlation between the predicted and true classifications. A higher MCC value (closer to 1) reflects a better prediction. As the value of this method depends on the detection threshold, we report the maximum MCC value over all thresholds.

\begin{equation}
    MCC = \frac{TP\cdot TN - FP \cdot FN}{\sqrt{(TP+FP)(TP+FN)(TN+FP)(TN+FN)}}
    \label{eq:mcc}
\end{equation}

In Fig.~\ref{fig:evaluation-diff10}, the MCC values are given for all three models on the test composite dataset \emph{without} recompression, for a range of QF pairs (each with a difference of 10). We can see that better performance is achieved for higher QF pairs, regardless of the model. The drop in performance for the highest QF pair using all three models can be explained by the fact that a QF of 100 was not included during training. When comparing HighQF and WideQF, we observe that including lower QFs during training does not significantly boost the performance of low QF pairs. In contrast, it decreases the performance in higher QF pairs. Therefore, we recommend using only relatively high QFs (e.g., 50 and higher) during training, which are also mostly used in practice.

In Fig.~\ref{fig:evaluation-recompress-high} and Fig.~\ref{fig:evaluation-recompress-rec}, the MCC values are given for HighQF and HighQFRec, respectively, on the \emph{recompressed} test composite dataset. We observe that HighQF is not robust against recompression at all: even when recompressing the composite images with a Rec. QF of 95 (i.e., resembling visually lossless compression), the performance drops significantly. In contrast, HighQFRec demonstrates robustness against recompression. Only when recompressing the composite images with a Rec. QF of 60 or lower, the performance is as low as HighQF. Therefore, we recommend including recompression in the training dataset. However, note that this comes at the cost of a slightly reduced performance on the test composite dataset \emph{without} recompression (see Fig.~\ref{fig:evaluation-diff10}).

\section{Conclusion}
This paper evaluated the impact of training data to create Comprint, an image forgery detection method. We conclude that having a training dataset with relatively high quality factors is sufficient, i.e., including relatively low quality factors during training does not contribute much.
Incorporating recompression during training, on the other hand, does have a positive effect on the performance against recompression attacks. Therefore, we advise including recompression during training to boost the Comprint's robustness in the wild.

\bibliographystyle{IEEEtran}
\bibliography{bib/IEEEabrv, bib/references-wo-doi}

\end{document}